\documentclass[fleqn,usenatbib]{mnras}

\usepackage{newtxtext,newtxmath}

\usepackage[T1]{fontenc}

\DeclareRobustCommand{\VAN}[3]{#2}
\let\VANthebibliography\thebibliography
\def\thebibliography{\DeclareRobustCommand{\VAN}[3]{##3}\VANthebibliography}

\usepackage{graphicx}	
\usepackage{amsmath}	
\usepackage{hyperref}

\title[Dwarf Galaxies at $z\sim2$]{The Origin of the Observed\\ Ly$\alpha$ EW Distribution of Dwarf Galaxies at $z\sim2$}

\author[C. Snapp-Kolas et al.]{
Christopher Snapp-Kolas,$^{1,{\href{https://orcid.org/0000-0002-9593-0053}{\includegraphics[height=0.3cm]{Images/orcidpic.pdf}}}}$\thanks{E-mail: csnappkolas@gmail.com}
Brian Siana,$^{1,{\href{https://orcid.org/0000-0002-4935-9511}{\includegraphics[height=0.3cm]{Images/orcidpic.pdf}}}}$
Timothy Gburek$^{1,{\href{https://orcid.org/0000-0002-7732-9205}{\includegraphics[height=0.3cm]{Images/orcidpic.pdf}}}}$
Anahita Alavi$^{2,{\href{https://orcid.org/0000-0002-8630-6435}{\includegraphics[height=0.3cm]{Images/orcidpic.pdf}}}}$
Najmeh Emami$^{3,{\href{https://orcid.org/0000-0003-2047-1689}{\includegraphics[height=0.3cm]{Images/orcidpic.pdf}}}}$
\newauthor Johan Richard$^{4,{\href{https://orcid.org/0000-0001-5492-1049}{\includegraphics[height=0.3cm]{Images/orcidpic.pdf}}}}$
Daniel P. Stark$^{5}$
Claudia Scarlata$^{6,{\href{https://orcid.org/0000-0002-9136-8876}{\includegraphics[height=0.3cm]{Images/orcidpic.pdf}}}}$
\\
$^{1}$Department of Physics \& Astronomy, University of California, Riverside, CA 92521, US\\
$^{2}$IPAC, California Institute of Technology, 1200 E. California Boulevard, Pasadena, CA 91125, USA\\
$^{3}$Minnesota Institute for Astrophysics, University of Minnesota, Minneapolis, MN, 55455, USA\\
$^{4}$Univ Lyon, Univ Lyon1, Ens de Lyon, CNRS, Centre de Recherche Astrophysique de Lyon UMR5574, F-69230, Saint-Genis-Laval, France\\
$^{5}$Steward Observatory, University of Arizona, 933 N Cherry Ave, Tucson, AZ 85721, USA\\
$^{6}$Minnesota Institute for Astrophysics, University of Minnesota, Minneapolis, MN, 55455, USA
}

\date{Accepted XXX. Received YYY; in original form ZZZ}

\pubyear{2022}

\begin{document}
\label{firstpage}
\pagerange{\pageref{firstpage}--\pageref{lastpage}}
\maketitle

\begin{abstract}
    We present a rest-UV selected sample of 32 lensed galaxies at $z\sim 2$ observed with joint Keck/LRIS rest-UV and Keck/MOSFIRE rest-optical spectra behind the clusters Abell 1689,  MACS J0717, and MACS J1149. The sample pushes towards the faintest UV luminosities observed ($-19 \le {\rm M_{\rm UV}} \le -17$) at this redshift. The fraction of dwarf galaxies identified as Ly$\alpha$ emitters ($\rm EW \ge 20\ \rm\AA$) is ${\rm X_{\rm LAE}}=25^{+15}_{-10}\%$. We use the Balmer lines and UV continuum to estimate the \textit{intrinsic} EW allowing us to distinguish the effects of the ionizing spectra and escape fraction on the observed EW distribution. Fainter galaxies (M$_{\rm UV} > -19$) show larger intrinsic EWs and escape fractions than brighter galaxies. Only galaxies with intrinsic EWs greater than $40\ \rm\AA$ have escape fractions larger than 0.05. We find an anti-correlation between the escape fraction and $\rm A_V$ as well as UV spectral slope. The volumetric escape fraction of our sample is $f_{\rm esc}^{\rm Ly\alpha} = 4.59^{+2.0}_{-1.4}\%$ in agreement with measurements found elsewhere in the literature. About half of the total integrated Ly$\alpha$ luminosity density comes from galaxies with ${\rm EW}_{\rm obs}>20$ \AA. 
\end{abstract}




\section{Introduction}
More than five decades ago, Ly$\alpha$ was proposed as a tracer of early stage star forming galaxies \citep{Partridge1967a}. Since then Ly$\alpha$ has been used to study star-forming galaxies locally and at the highest redshifts observed \citep[i.e.][]{Shapley2003,Jung2020}. It has been used to study reionization \citep{Bolton2013,Mesinger2015,Mason2018a,Matthee2021}, the interstellar medium (ISM) \citep{Du2021}, the circumgalactic medium \citep{Matsuda2012MNRAS.425..878M,Hayes2014a}, and has confirmed the redshifts of the most distant star-forming galaxies \citep{Caruana2014,Hoag2019b,Endsley2020,Jung2020}. 

Ly$\alpha$ is a resonant line and has a large cross section. Therefore, it will be scattered even by low column density gas. This makes it a great tracer of gas distributions, but can result in difficulties in interpreting observations. Furthermore, Ly$\alpha$ can be easily absorbed by dust in the interstellar medium (ISM) reducing the strength of the line. Ly$\alpha$ observations are even more difficult when we try to observe galaxies during or before the epoch of reionization where there are large amounts of neutral hydrogen in the intergalactic medium (IGM) which can scatter whatever Ly$\alpha$ manages to escape. Each of these effects can drastically hamper the observability of Ly$\alpha$ and have strong effects on the distribution of observed Ly$\alpha$ equivalent widths (EWs) \citep{Stark2010a,Stark2011a}.
In fact, at high redshift the attenuation of Ly$\alpha$ can be a useful probe of the neutral fraction of hydrogen in the intergalactic medium. But in order to measure the attenuation, one must first know the \textit{intrinsic} distribution of EW$_{\rm Ly\alpha}$ to compare with the observed distribution. The evolution of the EW$_{\rm Ly\alpha}$ distribution has been used to study the end of reionization and is well understood from z$=3-6$ and for "bright" galaxies with absolute UV magnitudes M$_{\rm UV}<-19$ \citep{Stark2010a,Stark2011a,Pentericci2011a,Curtis-Lake2012a,Ono2012a,Schenker2012,Cassata2015b,DeBarros2017a,ArrabalHaro2018,Caruana2018,Fuller2020,Delavieuville2020A&A...644A..39D,Kusakabe2020a,Zhang2021}. 
At $z=2$ there has been some study of the brighter star-forming galaxies ($\rm M_{\rm UV}<-19$) \citep{Reddy2009,Cassata2015b,Hathi2016,Du2021,Zhang2021}, but little work has been done to study the fainter population ($\rm M_{\rm UV}>-19$). However,  according to the luminosity functions of \citet{Alavi2016}, \citet{Konno2016}, and \citet{Bouwens2022arXiv220511526B} the faint galaxy population is larger at $z=2$ and becomes more numerous with redshift. Therefore, if we want to understand reionization we need more information on the faint galaxy population, and $z=2$ galaxies can help to constrain the EW$_{\rm Ly\alpha}$ distribution at higher redshift where observation of Ly$\alpha$ is more difficult. 

Ly$\alpha$ studies will either select galaxies via their rest-UV continuum luminosity density (via broadband imaging), or their Ly$\alpha$ line emission \citep[via narrowband imaging,][]{Berry2012b,Konno2016,Sobral2017,Sobral2019c,Hashimoto2017}.  Narrow-band selected samples will be biased towards Ly$\alpha$ emitters (LAEs) as it requires that the line be significantly detected in emission.
Alternatively, narrow-band studies of Ly$\alpha$ often select on other emission lines in the optical (such as [OIII]5007) \citep[i.e.][]{Ciardullo2014,Trainor2015f,Weiss2021}. \citet{erb2016ApJ...830...52E} show that [OIII]5007-selected galaxies are biased towards LAEs by comparing the fraction of LAEs (X$_{\rm LAE}$) in their [OIII]5007 selected sample with a sample of UV-selected galaxies. 80\% are found to be LAEs in their [OIII]5007 sample while only 9\% are found to be LAEs in their UV selected sample. This is further confirmed in \citet{Trainor2019,Weiss2021}, indicating that samples selected on nebular emission lines are biased toward Ly$\alpha$ emission. A sample chosen irrespective of the strength of nebular emission lines would be useful for understanding the full distribution of Ly$\alpha$ EWs. With an unbiased sample, it is possible to disentangle the effects of the stellar population (e.g. age and metallicity) and that of the gas and dust on the observed EW distribution. 

Broad-band surveys, which select galaxies independent of emission line strength, are able to probe down to ${\rm R}\sim 30$, but the followup spectroscopy necessary for accurate escape fraction and EW analyses limit the observations to $\rm R<25.5$ \citep[i.e][]{Reddy2008} in order to obtain sufficient S/N in the continuum. 

The Ly$\alpha$ EW distribution is dependent on the escape fraction of Ly$\alpha$ photons. Past studies have measured the Ly$\alpha$ escape fraction by comparing the Ly$\alpha$ luminosity density with the Ha or UV continuum luminosity densities, obtained by integrating luminosity functions \citep[][]{Hayes2010,Ciardullo2014,Konno2016}. However, comparing luminosity densities does not allow for analysis of individual galaxy escape fractions or correlations with other properties of the population. Others make use of narrowband imaging of Ly$\alpha$ and one of the Balmer lines \citep{Trainor2015f} to make direct measurements of the Ly$\alpha$ escape fraction. Narrow-band imaging is biased towards high Ly$\alpha$ EW galaxies, and therefore does not represent the general population of star-forming galaxies as evidenced by the rather high average Ly$\alpha$ escape fractions of 30\% \citep{Kornei2010,Blanc2011,Wardlow2014,Trainor2015f} which differs greatly from other measures with more complete samples \citep[i.e.$\;5.3\%$,][]{Hayes2010}. Some try to model the escape fraction as a function of Ly$\alpha$ EW \citep[i.e.][]{Sobral2019c}, but this method must assume a star formation history and intrinsic ionizing continuum which are uncertain for low-mass galaxies. A direct measurement from a UV-selected sample with spectroscopic followup avoids these uncertainties. 

What is needed is a measure of Ly$\alpha$ emission properties of a UV-selected sample of faint galaxies. We address this by selecting galaxies with photometric redshifts determined from HST photometry described in \citet{Alavi2014} and \citet{Alavi2016}. We did not select on emission line flux and therefore the sample is unbiased towards Ly$\alpha$ emitters. Furthermore, by using gravitational lensing, we can study much fainter galaxies at $z=2$. The goals of this paper are:

\begin{itemize}
    \item Measure the Ly$\alpha$ EW distribution and determine its dependence on the properties of the ionizing sources and the Ly$\alpha$ escape fraction.
    \item Calculate the volumetric escape fraction of dwarf galaxies
    \item Identify trends in the EW distribution with absolute UV magnitude and redshift
\end{itemize}

    The remainder of the paper is organized as follows. In $\S$\ref{Observations}, we discuss the observations, data reduction, and sample selection. In $\S$\ref{Methods}, we discuss fits to continua and emission lines, SED fitting, magnification, and dust estimates. In $\S$\ref{Results}, we present our measured values of the EW distribution and the Ly$\alpha$ escape fraction. In $\S$\ref{discussion}, we discuss the implications of our measurements. In $\S$\ref{Summary}, we summarize the work. We adopt a $\Lambda$CDM cosmology with $\Omega_m$ = 0.3, $\Omega_{\Lambda} = 0.7$, and $h = 0.7$ throughout the paper and all magnitudes are in the AB system.
\section{Observations, Data Reduction, and Sample Selection}
\label{Observations}
\subsection{Observations \& Data Reduction}
Our sample consists of dwarf galaxies behind lensing clusters that have deep HST data and are accessible to the Keck Observatory (Abell 1689, MACS J1149, MACS J0717). Photometric redshifts were determined from HST photometry, as described in \citet{Alavi2014, Alavi2016}. Our selection criteria were photometric redshifts of $1.5<z<3.5$ and visual magnitudes brighter than m$_{\rm F625W} < 26.3$. Slit allocation preference was given to galaxies with high magnification and photometric redshifts where the rest-optical lines are accessible in the YJHK atmospheric bands. In order to study physical properties (i.e. metallicity, $f_{\rm esc}$, ionization parameter, etc.) of these dwarf galaxies we obtained  Keck/MOSFIRE spectra as described in \citet{Gburek2019}. We also obtained Keck/LRIS optical spectra, for which the details are listed in Table 1. We observe eleven masks in three clusters with LRIS. Exposure times varied greatly (from 4500s-18000s) depending on the conditions and priority of objects in the mask, and seeing was typically $\sim 1''$. We decreased read noise by binning by 2 in the spectral direction of the CCD. 

\begin{table}[h]
	\centering
	\caption{List of data acquired with Keck/LRIS. In order from left to right we have the mask name used, the number of spectra obtained in each mask, the exposure time in seconds, the modified julian date (MJD), and the seeing in arcseconds}
	\label{tab:example_table}
	\begin{tabular}{lcccc} 
		\hline
		Mask & Spectra & Exposure Time & MJD & Seeing\\
		(Name) & (Number) & (s) & & (arcsec)\\
		\hline
        A1689$\_$1 & 15 & 18000 & 55325 & 0.9 \\ 
        A1689$\_$3 & 16 & 12286 & 55981 & 0.9 \\ 
        Macsj0717 & 11 & 12600 & 55981 & 1.1 \\
        Macsj1149$\_$1 & 16 & 5400 & 57042 & 1.1 \\
        Macsj0717${\_}2$ & 8 & 9000 & 57042 & 0.7 \\
        Macsj0717$\_$1 & 8 & 9000 & 57042 & 0.9 \\
        A1689$\_$4 & 10 & 9000 & 57042 & 1.4 \\
        Macsj1149$\_$1\_2  & 13 & 5400 & 57043 & 0.8 \\
        A1689$\_z1\_$1 & 12 & 12000 & 57043 & 0.8 \\
        Macsj1149$\_$2 & 5 & 4500 & 57398 & 1.0 \\
        Macsj0717$\_$3 & 8 & 6720 & 57398 & 0.7 \\
        Macsj1149$\_$3 & 11 & 4860 & 57870 & 1.5 \\
        A1689$\_$6 & 10 & 9600 & 57870 & 1.0 \\
		\hline
	\end{tabular}
\end{table}

The data were reduced using a modified version of the PypeIt v1.x reduction pipeline \citep{pypeit:joss_arXiv}. This pipeline performs bias subtraction,  flat fielding, cosmic ray rejection, wavelength calibration, sky subtraction, and extracts the 1D spectra. The 1D spectra are extracted using an optimal extraction b-spline fitting to the object profile along the spatial direction. The 1D spectra are then flux calibrated using the spectrum of a standard star. The 1D spectra from each frame are then combined using a weighted mean algorithm in PypeIt. We perform a final absolute flux normalization by scaling the spectra of compact bright continuum sources to the {\it Hubble} photometry. We take the median (per mask) of these and scale the remaining spectra according to this median. Most of our galaxies are faint and therefore may not have their continua well-detected in the 1D spectra, so the use of a median correction from bright continuum sources is necessary for the absolute calibration. 

We correct for slit losses by convolving the Hubble images with a Guassian such that point sources in the final images would have a Guassian FWHM equal to the seeing. We then measure the total flux and the flux within the slit. We use the F435W and F475W bands of Hubble for the Abell 1689 and MACS J0717/MACS J1149 masks respectively. We apply this correction to Ly$\alpha$, to obtain an estimate of the total flux, but refrain from applying the correction when displaying spectra. 

 When we have spectra of the same galaxy in different masks, we combine them using an inverse variance weighted mean. Some of our galaxies are multiply imaged, but combining multiple images involves demagnifying the spectra before combining and results in a spectrum of arbitrary magnification. Furthermore, the uncertainties in the magnification of high magnification galaxies would result in large uncertainties in the flux with little gain in S/N. Therefore, we remove highly magnified images for the multiply imaged galaxies. 

\begin{figure}
    \centering
    \includegraphics[height=17cm]{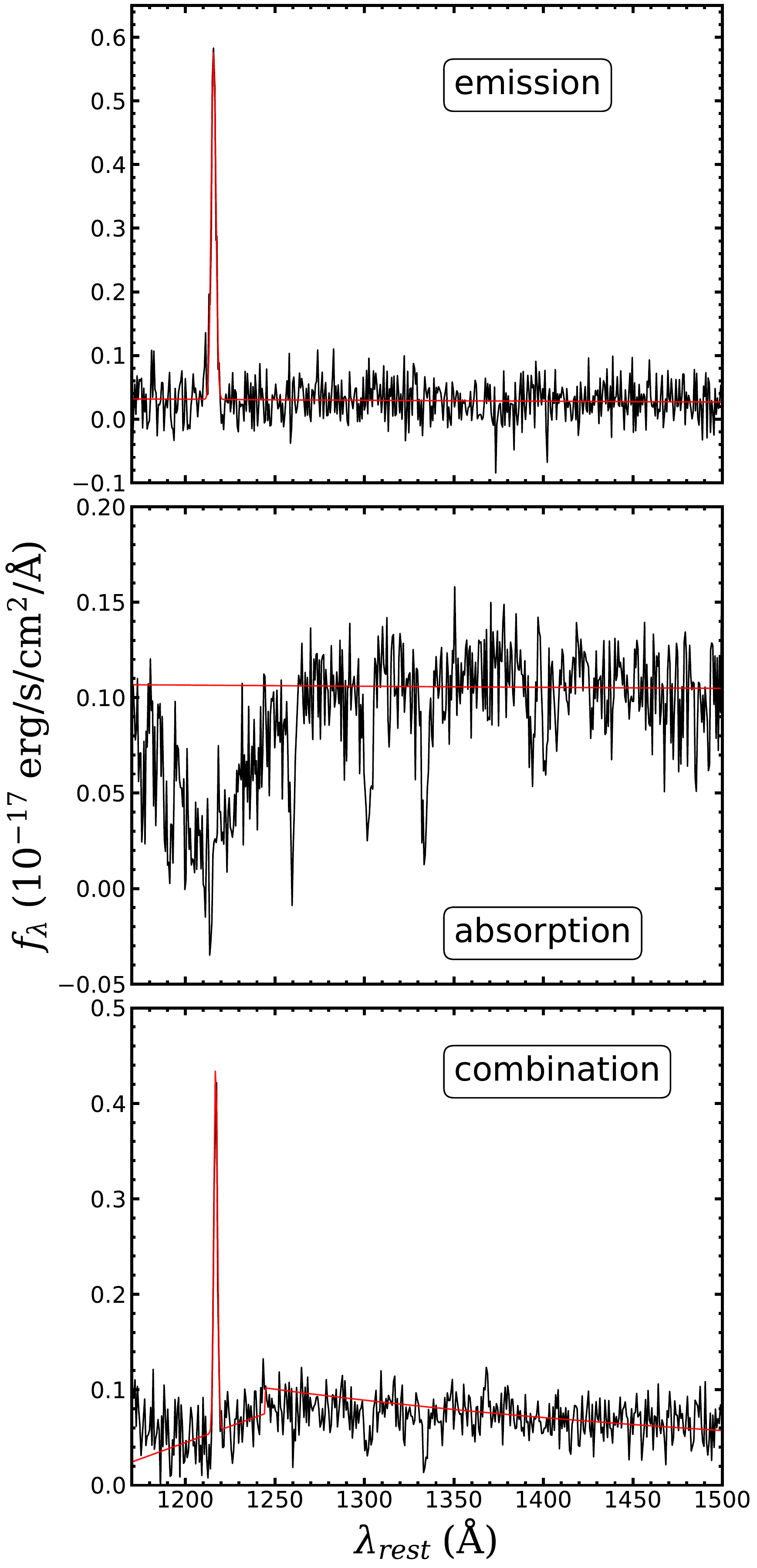}
    \caption{Three example spectra from our sample of galaxies. The red line is the fit for each spectrum. The fitting method is described in section \ref{Methods}. Top: An example of an ``emission" spectrum where we have Ly$\alpha$ in emission only. Middle: An example of an ``absorption" spectrum. Bottom: An example of a ``combination" spectrum where we see absorption and emission around Ly$\alpha$}
    \label{fig:spectra}
\end{figure}

The LRIS slit masks use box slits to align the masks on bright stars in the field. Occasionally, the spectra of these alignment stars will contaminate nearby slits. As a result some of the slits near the alignment stars are rejected due to poor sky subtraction. After accounting for slits affected by nearby box slits, our final count of 1D spectra is 127. The clusters are about 3'x3' which takes up about one half of the LRIS detector. Because of this we have about half the number of spectra typically expected from an LRIS mask. We estimate the spectral resolution of our spectra given the slit width of our observations is $\sim$1"  with a plate scale of 0.135"/pix, however we have binned along the wavelength direction such that the effective plate scale is 0.27"/pix. The wavelength spacing is $\sim 2.18 \; \rm \AA\;\rm pix^{-1}$. So the resolution of our observations is $\rm R\sim 500$ or $\Delta v \sim 600 \;\rm km\;\rm s^{-1}$. We show some example 1D spectra in figure \ref{fig:spectra}.

\subsection{Sub-Sample Selection}
The remainder of the paper focuses on an analysis of X$_{\rm LAE}$ and the escape fraction of Ly$\alpha$ photons ($f_{\rm esc}$). In order to measure X$_{\rm LAE}$ in an unbiased manner we need to select a sub-sample of galaxies for which we can identify galaxies with no Ly$\alpha$ emission. That is, a spectroscopic redshift can be measured independent of detection of the Ly$\alpha$ line. We call this sub-sample the X$_{\rm LAE}$ sample. In order to measure $f_{\rm esc}$ we must be able to measure H$\alpha$. This is a more stringent requirement which eliminates a greater number of galaxies from the sample and would result in greater uncertainty in X$_{\rm LAE}$ were we to impose this condition on all of our data. Therefore, we choose to create an additional sub-sample which we call the H$\alpha$ sample.

\subsubsection{X$_{\rm LAE}$ sample}
\begin{figure}
    \centering
    \includegraphics[height=7cm]{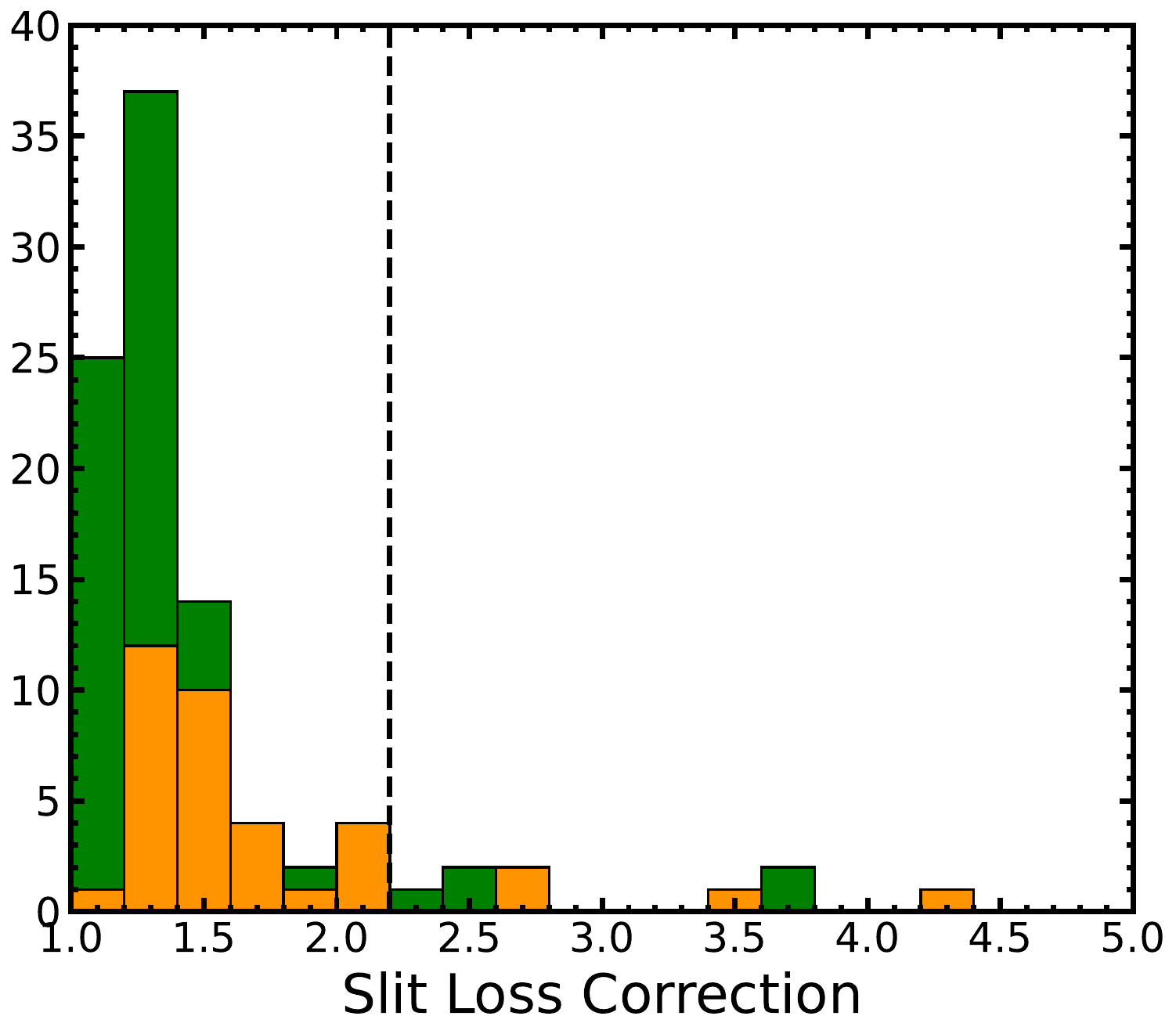}
    \caption{The distribution of slit loss corrections to our H$\alpha$ and Ly$\alpha$ spectra. The dashed line at 2.2 denotes the upper limit we set for including galaxies in our sample.}
    \label{fig:slit}
\end{figure}

We choose a sub-sample of galaxies from our sample of 89 in order to avoid biases in our Ly$\alpha$ equivalent width distribution. We account for biases from magnification, slit-losses, and possible blending of sources. We mimic the sub-sample selection methods of \citet{Emami2020}, but with values derived from our sample as follows: 
\begin{itemize}

    \item We require secure redshifts for all the galaxies in this sample so that galaxies with no Ly$\alpha$ or Ly$\alpha$ in absorption can be identified. Therefore, only galaxies with confirmed redshifts from our MOSFIRE data are kept. These redshifts are primarily determined by H$\alpha$ and [OIII]$\lambda\lambda5007$.
    
    \item We remove galaxies which do not have spectroscopic coverage of Ly$\alpha$. Atmospheric absorption at $\lambda < 3200$ \AA\ limits our analysis to galaxies at $z\gtrsim1.6$.
    
    \item We remove galaxies for which there is possible confusion/blending with nearby objects in the slit. These objects could have emission lines from one galaxy and continuum from another, affecting the equivalent width measurements.
    
    \item We remove galaxies with large magnification, as they could suffer from differential magnification across the galaxy. This could introduce a bias into our sample which is selected on UV luminosity density. We therefore remove galaxies which have magnification $\mu > 30$ for Abell 1689 and $\mu>15$ for MACS J0717 and MACS J1149.
    
    \item Finally, we remove galaxies with large slit loss correction in either LRIS or MOSFIRE.  If a large fraction of the galaxy flux is outside of a slit, then it is not clear that a slitloss-corrected flux will reflect the true spectrum. Most of the galaxies have slit losses $<2.2$ and we thus set this as our upper limit. The slit loss distributions are shown in figure
    \ref{fig:slit}. 
\end{itemize}

After this cleaning of the sample we obtain a X$_{\rm LAE}$ sample of 32 galaxies with a mean redshift $\langle z \rangle = 2.27$.

\subsubsection{H$\alpha$ sample}
\begin{figure}
    \centering
    \includegraphics[height=7cm]{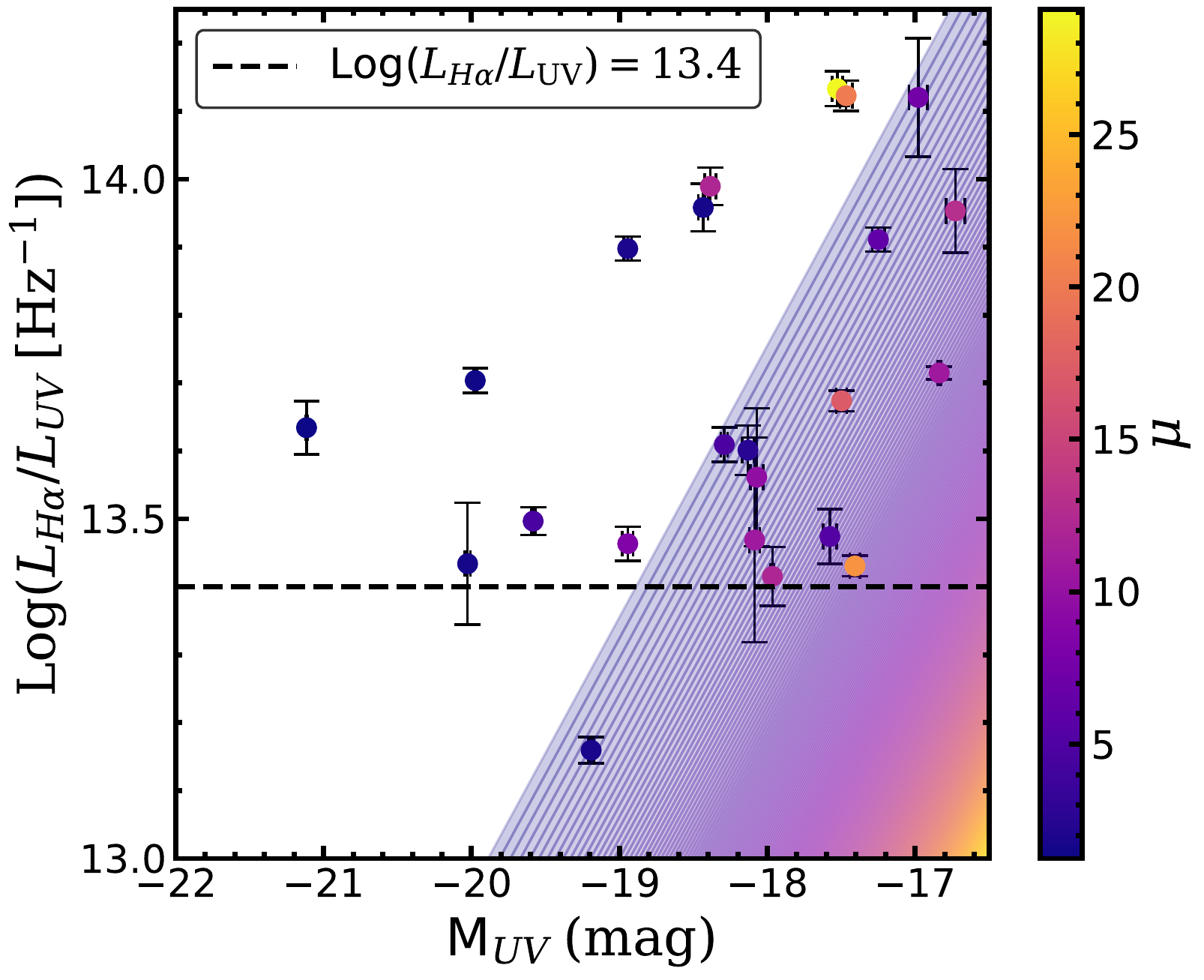}
    \caption{Plotted is the $\rm log(L_{H\alpha}/L_{UV})$ vs. absolute UV magnitude. Points are color coded according to each galaxy's magnification. The black dashed line at $\rm log(L_{H\alpha}/L_{UV}) = 13.4$ denotes our completeness limit for the sample. The angled colored space demonstrates the 3$\sigma$ H$\alpha$ sensitivity limit of our detector color coded according to magnification}
    \label{fig:LhLUV}
\end{figure}

In order to get a measurement of the Ly$\alpha$ escape fraction, H$\alpha$ must be observable. Because of this we create a second sub-sample of galaxies from the X$_{\rm LAE}$ sub-sample for which H$\alpha$ could have been observed, which limits the redshift range to $z\lesssim2.6$. Additionally, The rest-UV HST data are very deep and the KECK H$\alpha$ data are shallower which could bias us against galaxies that are faint in H$\alpha$. To check whether we are excluding galaxies that are too faint in H$\alpha$ we observe that our bright M$_{\rm UV}$ sample lies almost entirely above log(L$_{\rm H\alpha}$/L$_{\rm UV}$)$\sim 13.4$, as shown in figure \ref{fig:LhLUV}, and therefore set this as our completeness limit for the sample. We take the median H$\alpha = 4.3\times 10^{-18}\;\rm erg\;s^{-1}\;cm^{-2}$ line flux error multiplied by 3 as our 3$\sigma$ limit for detection of H$\alpha$. We then check our galaxies magnification and absolute UV magnitude and ask whether the galaxy would have been observable at log(L$_{\rm H\alpha}$/L$_{\rm UV}$)$\sim 13.4$. If the galaxy would not have been observable at $3\sigma$ at our completeness limit it was then removed. No galaxy fell below this criterion and therefore none were removed as shown in figure \ref{fig:LhLUV}. We split our sample at M$_{\rm UV}=-19$ and measure the mean and error on the mean of the bright and faint samples to be log(L$_{\rm H\alpha}$/L$_{\rm UV}$)$= 13.8\pm0.003$ and log(L$_{\rm H\alpha}$/L$_{\rm UV}$)$= 13.5\pm0.002$ respectively.  While all but one of our galaxies have log(L$_{\rm H\alpha}$/L$_{\rm UV}$)$> 13.4$, 11 of the 18 galaxies fainter than M$_{\rm UV} = -19$ could have been observed below this line, given their magnifications. This suggests that there are indeed very few galaxies with log(L$_{\rm H\alpha}$/L$_{\rm UV}$)$<13.4$ and that our measured increase of log(L$_{\rm H\alpha}$/L$_{\rm UV}$) for faint galaxies is real, and not due to a bias against galaxies with low log(L$_{\rm H\alpha}$/L$_{\rm UV}$). Our final H$\alpha$ sample after all of these considerations is 23 galaxies.


\section{Methods}
\label{Methods}
\subsection{H$\alpha$ fits}
The fits to the H$\alpha$ line were performed in a manner similar to \citet{Gburek2019} on the 1D extracted MOSFIRE spectra. We briefly review the procedure here. The spectra were fit using an MCMC sampler algorithm \emph{emcee} \citep{Foreman_Mackey_2013}. Each of the J,H, and K bands were fit separately and the emission lines were parameterized with a Gaussian and a linear continuum fit. The redshift was taken to be the weighted average of the values obtained between bands with emission lines present. The flux of the H$\alpha$ line was taken from the Gaussian parameterization.

\subsection{Ly$\alpha$ fits}
 We identified four types of spectra with respect to the Ly$\alpha$ line in our dataset. Adopting the naming conventions of \citet{Kornei2010} we observe Ly$\alpha$ in 'emission', 'absorption', 'combination', and 'noise.' We measure the EWs of these spectra using the conventions of \citet{Du2018} with some variation. For each type of galaxy we set the bounds of numerical integration as follows: 
\begin{itemize}
    \item For emission line galaxies we set the bounds to be where the emission line met the continuum 
    \item For combination and absorption line galaxies we set an upper limit to the width at 45 \AA\ rest frame. The long-wavelength bound was set at the location where the absorption profile met the continuum fit. The short-wavelength bound was set at the blue end of the absorption wing or redder if this resulted in a width greater than 45 \AA\ rest-frame.
    \item For noise spectra we have no information on the Ly$\alpha$ profile. Therefore, we use boundaries derived by \citet{Kornei2010} at 1199.9 \AA\ and 1228.8 \AA\ rest-frame for the short- and long-wavelength bounds respectively. Where the redshift is determined from the rest-optical emission lines.
\end{itemize}

 In addition, we performed a parameterized fit of the Ly$\alpha$ profile (for emission and combination spectra) and continuum of each spectrum for our galaxy sample. The fitting procedure is performed using a Markov chain Monte Carlo code from pymc3. The wavelength range used for fitting varied based on redshift as the short-wavelength bound was set to just blueward of the Ly$\alpha$ emission line. The long-wavelength bound was set to be at $5450$ \AA\ in the observed frame, as the transmission decreases at longer wavelengths due to the dichroic at 5600$\rm\AA$. Typical emission and absorption lines (SiII$\lambda\lambda 1260$ \AA, OI+SiII$\lambda\lambda 1303$ \AA, CII$\lambda\lambda 1334$ \AA, SiIV$\lambda\lambda 1393$ \AA, CIV$\lambda\lambda 1549$ \AA, HeII$\lambda\lambda 1640$ \AA, OIII$\lambda\lambda 1666$ AA, CIII$\lambda\lambda 1909$ \AA, and OIV$\lambda\lambda 1343$ \AA) were masked when fitting each spectrum. We set the width for these masks to be $800{\;\rm km\;s^{-1}}$ except for the CIV$\lambda\lambda 1549$ \AA\ line which we set to have an asymmetric width of $1400{\;\rm km\;s^{-1}}$ with the blue side of the line extending to $-1000{\;\rm km\;s^{-1}}$ because of the P-Cygni profile that CIV$\lambda\lambda 1549$ \AA\ tends to take. This process was repeated for all galaxies observed. The `emission' spectra are modelled simultaneously with a power law continuum and a Gaussian emission line to handle correlations between parameters, 
 
\begin{equation}
f_{\lambda} =  A_0 e^{-\frac{(\lambda - \lambda_0)^2}{2\sigma^2}} + B_0 (\frac{\lambda}{\lambda_c})^\beta 
\end{equation}
 For the 'combination' spectra we add in a first order approximation to the Ly$\alpha$ absorption profile, 
 \begin{equation}
 f_{\lambda} = \left\{
 \begin{array}{ll}
      A_0 e^{-\frac{(\lambda - \lambda_0)^2}{2\sigma^2}} + a\lambda+b, & \lambda < \lambda_s \\
      B_0 (\frac{\lambda}{\lambda_c})^\beta, & \lambda > \lambda_s
 \end{array}
 \right.
 \end{equation}
 
 Where $\lambda_s$ is the wavelength where the linear approximation to the Voigt profile meets the UV continuum.
 The remaining galaxies are parameterized with a power law continuum redward of Ly$\alpha$ (identified either from damped Ly$\alpha$ absorption or from the redshift estimate obtained with Keck/MOSFIRE),
 \begin{equation}
    f_{\lambda} = A_0 (\frac{\lambda}{\lambda_0})^\beta
 \end{equation}

We use an upper limit of FWHM = 4.9 \AA\ rest-frame to obtain upper limits on the flux from the `absorption' and `noise' spectra. The value of 4.9 \AA\ is derived from the distribution of 49 FWHMs from the sample for which we have Ly$\alpha$ in emission. 4.9 \AA\ is greater than all but two of 49 (4\%) of the FWHMs measured and therefore serves as an upper limit on the FWHM. We determine the Ly$\alpha$ EW and flux by numerical integration around Ly$\alpha$ using the methods of \citet{Du2021} and then divide by the continuum level extrapolated to rest-frame 1216 \AA\ to determine the EW. 

\subsection{Magnification}

Thanks to lensing, the signal from faint galaxies is magnified and we are able to extend spectroscopic studies to fainter magnitudes at $z\sim2$ in just a few hours of Keck/LRIS and Keck/MOSFIRE time (see table \ref{table1}). Our galaxies are observed behind three lensing clusters. In the X$_{\rm LAE}$ and H$\alpha$ sub-samples the magnifications span $1.25 \le \mu \le 29.12$. The median magnifications are  $\langle\mu\rangle_{\rm median} = 7.11$ and $\langle\mu\rangle_{\rm median} = 7.45$ respectively. It is important to properly model the lensing in order to accurately measure the UV luminosity of the galaxies. \citet{Alavi2016} details the lens models used to measure the magnifications of the galaxies behind MACS J0717, MACS J1149, and Abell 1689. The models considered are constrained by the location and redshift of known multiply-imaged sources. We use the models produced by the Clusters As Telescopes (CATS) collaboration\footnote{\url{https://archive.stsci.edu/prepds/frontier/lensmodels/}}. As \citet{Alavi2016} states, we use the models of \citet{Limousin2007ApJ...668..643L}, \citet{Limousin2016} and \citet{Jauzac2016} for Abell1689, MACS J0717, and MACS J1149 respectively. These models are derived using \texttt{LENSTOOL} \footnote{\url{https://projets.lam.fr/projects/lenstool/wiki}} \citep{Jullo_2007}.

\subsection{SED fits}
We fit spectral energy distributions (SEDs) to our Hubble photometry in the near-UV, optical, and near-IR. Before SED-fitting we subtract off contributions from nebular emission lines using our slit-loss corrected spectra. We add an additional 3\% flux error to all of our bands to account for systematic errors in our photometry \citep{Alavi2016}. Using the code \texttt{FAST}\footnote{\url{https://w.astro.berkeley.edu/~mariska/FAST.html}} \citep{Kriek2009}, we fit \citet{Bruzual2003MNRAS.344.1000B} stellar population synthesis models to the emission-line subtracted photometry under the following assumptions:
\begin{itemize}
    \item Constant star formation histories (SFHs)
    \item A \citet{Chabrier2003} initial mass function (IMF)
    \item Stellar metallicities of either 0.2 $Z_{\odot}$ or 0.4 $Z_{\odot}$
    \item A \citet[SMC][]{Gordon2003} dust attenuation curve
\end{itemize}
Galaxy redshifts are set by the fit curves to the spectroscopy.

\subsection{Dust Correction}
In our analysis of the H$\alpha$ sample we need to correct the H$\alpha$ line and the UV continuum at 1700 \AA\ for dust attenuation. We use the SMC curve of \citet{Gordon2003} and A$\rm _V$ estimates from our SED fitting to estimate the attenuation of the UV continuum. The correction for H$\alpha$ is estimated by direct measurement of the Balmer decrement (assuming Case-B recombination). However, many of our galaxies have Balmer lines that are either too faint or obscured by sky emission and we are unable to directly measure the Balmer decrement of each galaxy. We observe that the Balmer decrement is well determined for three of our five brightest galaxies , while our fainter galaxies are less certain. Therefore, We split our H$\alpha$ sample into bright (M$_{\rm UV}<-19$) and faint (M$_{\rm UV}>-19$) sub-samples for which we estimate A$\rm _V$. For the faint sub-sample we normalize the spectra by H$\alpha$ and perform a median stack of the galaxies. We use a \citet{cardelli1989} curve and measure $\langle{\rm A_V}\rangle = 0.54$ and apply this to all of our dwarf galaxies. Most of our reported results will be about the dwarf galaxies for which we use this average attenuation value. We recognize that there will be significant scatter around this value. Nonetheless, the correction is small, a factor of 1.5, and does not affect the average trends reported in this paper. For our more massive galaxies we measure A$\rm _V$ for three of the galaxies (${\rm A_V = 1.14,\ 0.88,\ and\ 0.80}$). We then use the mean ($\langle{\rm A_V}\rangle = 0.94$) on the remaining two galaxies for which we did not have H$_{\beta}\;S/N > 5$ and therefore could not reliably calculate the dust correction.

\section{Results}
\label{Results}
\begin{figure}
    \centering
    \includegraphics[height=7cm]{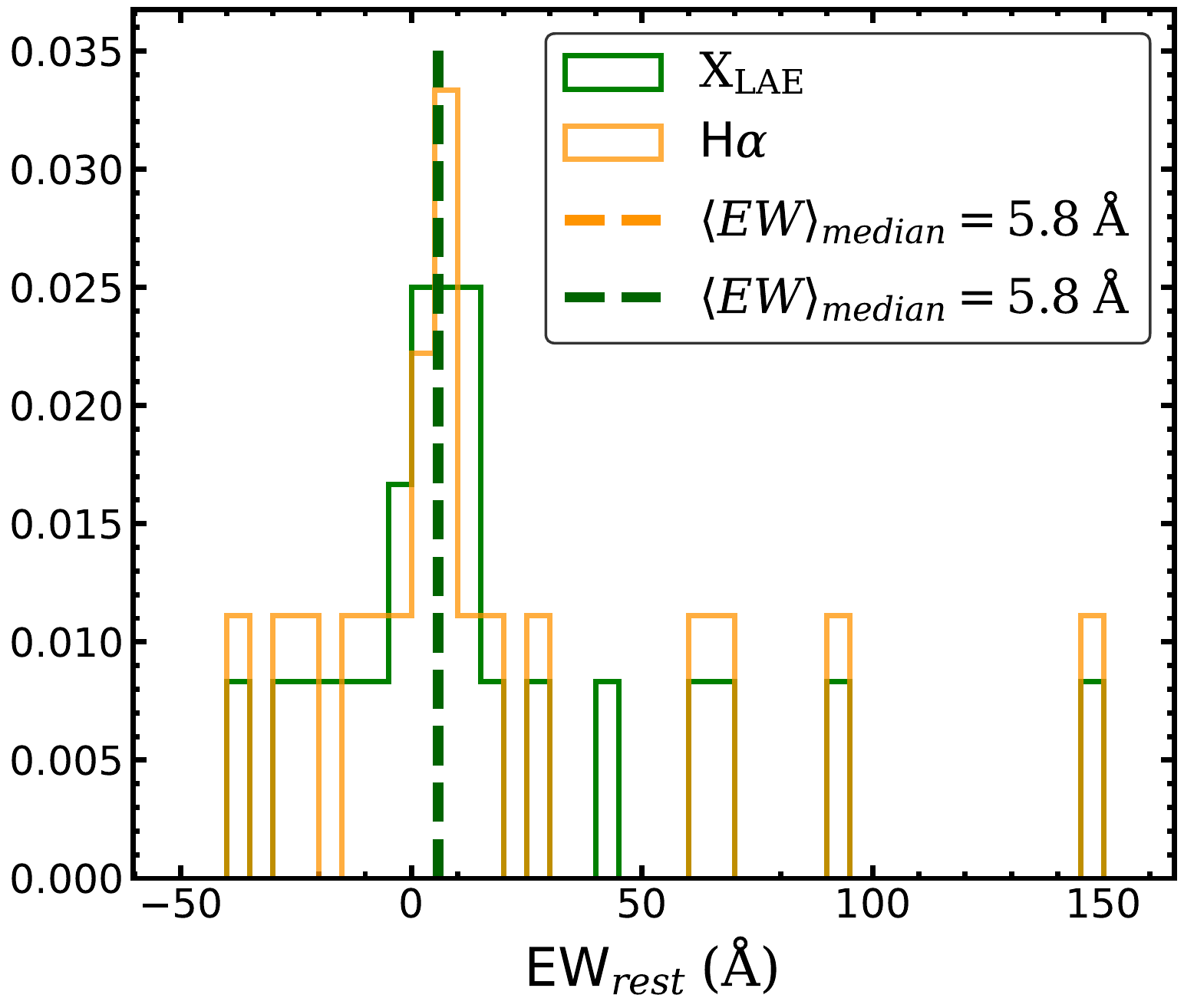}
    \caption{The green (orange) histogram is a normalized distribution of rest-frame EWs of our $\rm X_{LAE}$ (H$\alpha$) sample. The colored dashed vertical lines show the median EW of 5.8 (5.8) for the $\rm X_{LAE}$ (H$\alpha$) sample. The H$\alpha$ median (orange dashed line) is under the green dashed line. The Ly$\alpha$ distributions of both samples are consistent.}
    \label{fig:big_small_EW}
\end{figure}

As explained above, our larger $\rm X_{LAE}$ sample is used to measure the EW distribution and the smaller Ha sub-sample is used to measure the escape fraction. We compare our two samples in order to see if the two are similar and therefore both representative of dwarf galaxies at this redshift. We find that the median EWs are consistent with one another as shown in figure \ref{fig:big_small_EW}. The difference in mean redshift is also small ($\langle z_{\rm Ly\alpha}\rangle = 2.28$ and $\langle z_{\rm H\alpha}\rangle=2.23$). This suggests that the trends of the H$\alpha$ sample are likely representative of the $\rm X_{\rm LAE}$ sample as well. 

\subsection{Ly$\alpha$ EW distribution}
\begin{figure}
    \centering
    \includegraphics[height=7cm]{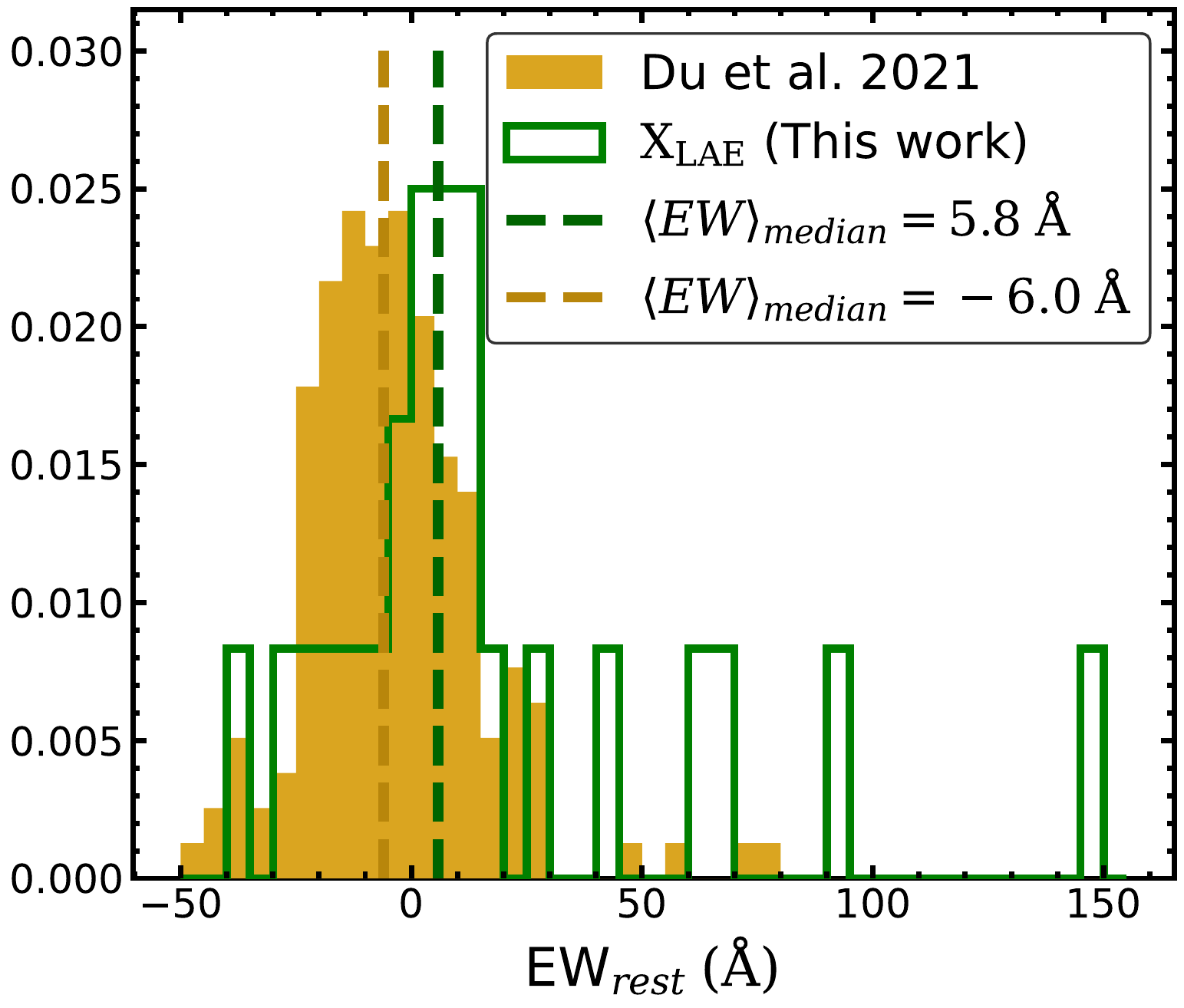}
    \caption{The green histogram is a normalized distribution of rest-frame EWs of our $\rm X_{LAE}$ sample. The gold histogram is a normalized distribution of rest-frame EWs of the brighter sample from \citet{Du2021}. The vertical dashed lines of corresponding color show the median Ly$\alpha$ EWs of each sample. the median EWs are $\langle EW \rangle_{\rm median}=5.8\AA$ and $\langle EW \rangle_{\rm median}=-6.0\AA$ respectively.}
    \label{fig:Du}
\end{figure}

We compare our EW distribution with that of the higher luminosity sample of \citet{Du2021} as shown in Figure \ref{fig:Du}. We find that our sample is skewed towards larger Ly$\alpha$ EWs. The median EW for the \citet{Du2021} sample is $-6.0$ \AA\ and the median for our sample is $5.8$ \AA. 

The cutoff for defining a LAE in the literature varies significantly depending on the study \citep[][etc.]{Stark2010a,Stark2011a,Caruana2018,Kusakabe2020a}. However, the most common cuts are $\rm EW > 20$ \AA, $EW > 25$ \AA, and $EW > 55$ \AA, so we will use these cuts as well. For these three cutoffs we measure X$_{\rm LAE}$ = $25^{+15}_{-10}\%$ (6/24), $25^{+15}_{-10}\%$ (6/24), and $17^{+13}_{-8}\%$ (4/24), respectively. We assume a Poisson distribution to estimate uncertainties on $\rm X_{LAE}$. X$_{\rm LAE}$ measurements of brighter galaxies are  $\sim11\%$ for $EW > 20$ \AA\ at $z\sim2$ (12\% from \citet{Reddy2008}, 11.1\% from \citet{Hathi2016}, and 10.7\% \citet{Du2021}). For brighter galaxies with $EW > 25$ \AA\ and $EW > 55$ \AA\ at $z\sim 2$ X$_{\rm LAE} = 10\%$ and $4\%$ respectively \citep{Cassata2015b}. These data suggest that galaxies fainter than M$_{\rm UV}=-19$ have greater numbers of LAEs in their population. We explore this in more detail in section \ref{Sec:XLAE}. 

\subsection{Escape Fraction}
In our H$\alpha$ sample for which we are able to measure the escape fraction of Ly$\alpha$ photons we assume a Case-B scenario for which the gas is taken to be optically thick for the Lyman series. The ratio of the flux of Ly$\alpha$ to the flux of H$\alpha$ under this assumption can vary depending on the temperature and electron density of the galaxy. We choose to use a ratio of 8.7 which is common in the literature \citep[i.e.][]{Matthee2016}. We then measure the escape fractions of individual galaxies as
\begin{equation}
    f_{\rm esc} = \frac{F_{\rm Ly\alpha}}{8.7\ F_{\rm H\alpha,cor}}
    \label{eq:3}
\end{equation}

where $\rm F_{Ly\alpha}$ is the Ly$\alpha$ flux and $\rm F_{H\alpha,cor}$ is the dust corrected H$\alpha$ flux. The mean of the individual $f_{\rm esc}$ measurements is $4.3^{+1.6}_{-1.1}\%$. We estimate the uncertainty in the mean via a bootstrap method. We perturb each measurement of the sample according to a Gaussian distribution with standard deviation set by the uncertainty of the measurement and then resample by randomly selecting with replacement from our perturbed sample and then calculating the mean of the escape fractions. This process is performed 100000 times. We then take the 16th and 84th percentiles of the 100,000 iterations to be the uncertainty on the mean escape fraction. 

However, this value may not be representative of the actual number of photons escaping into the intergalactic medium. The intrinsic luminosity of galaxies varies greatly and may correlate with Ly$\alpha$ escape fraction. Thus, the mean escape fraction may be different from the net output from all galaxies, which we refer to as the {\it volumetric} escape fraction. To calculate the volumetric escape fraction, we first need to determine the galaxies' intrinsic luminosity and, thus, de-magnify each of our galaxies. We then replace the fluxes in \ref{eq:3} with the sum of the respective luminosities. We measure a volumetric escape fraction of $4.59^{+2.0}_{-1.4}\%$. We estimate the uncertainty in a similar manner to the individual escape fraction estimate. We perturb each measurement of the sample according to a Gaussian distribution with standard deviation set by the uncertainty of the measurements, resample, and then calculate the volumetric escape fraction. We take the 16th and 84th percentiles of the 100,000 iterations to be the uncertainty on the volumetric escape fraction. This method includes contributions from uncertainties in the dust correction. This value agrees well with what is found in the literature ($5.3\%\pm3.8\%$ in \citet{Hayes2010}, $4.4^{+2.1}_{-1.2}\%$ in \citet{Ciardullo2014}, $5.1\%\pm 0.2\%$ in \citet{Sobral2017}, and $5.8^{+0.7}_{-0.5}\%$ in \citet{Weiss2021}). 

By the use of gravitational lensing we were able to directly measure a UV selected sample of dwarf galaxies at $z=2$. We are able to probe fainter H$\alpha$ and Ly$\alpha$ luminosities and have a greater sample of joint detections of H$\alpha$ and Ly$\alpha$ emitters than what was observed in \citet{Hayes2010}. Furthermore, our sample is not biased towards emitters as in \citet{Ciardullo2014}, \citet{Sobral2017}, and \citet{Weiss2021}. Nevertheless, the volumetric escape fraction in each study is consistent with our measurement, and all within 4\%-6\%.

\section{Discussion}
\label{discussion}
Here we seek to understand the origins of the observed EW distribution. Is the spread in observed EW primarily due to varying escape fractions, or are variations in stellar populations (in particular, starburst age), creating a large scatter in {\it intrinsic} EWs as well. 

\subsection{Intrinsic Ly$\alpha$ EW}

\begin{figure}
    \centering
    \includegraphics[height=7cm]{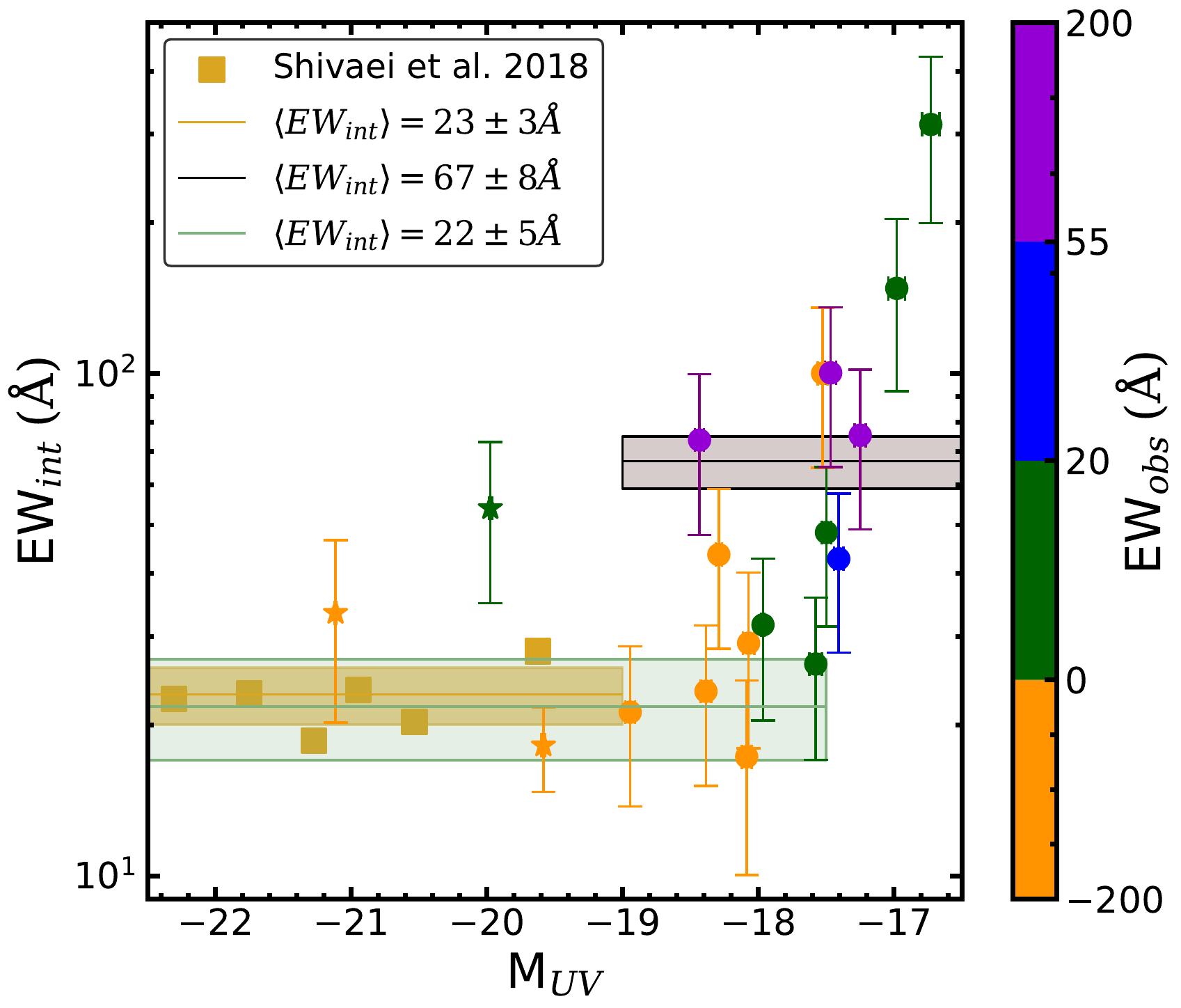}
    \caption{The \textit{intrinsic} equivalent width as a function of absolute UV magnitude (M$_{\rm UV}$). Points are color coded according to the \textit{observed} equivalent width (see color bar on the right) with blue and purple points representing galaxies observed as LAEs. The square gold points are average values of higher luminosity galaxies and are derived from the $\xi_{\rm ion}$ measurements of \citet{Shivaei2018}. Galaxies brighter than -19 in absolute UV magnitude are plotted with an x, while fainter galaxies are plotted with filled circles. We show the mean intrinsic EW of \citet{Shivaei2018} and uncertainty in the mean with a gold shaded region, and we show the mean intrinsic EW and uncertainty in the mean of our dwarf galaxy sample in black. Five galaxies did not have sufficient photometric coverage to perform SED fitting and are removed from any intrinsic EW analysis.}
    \label{fig:intMUV}
\end{figure}

To investigate influences from the stellar populations we make use of the intrinsic Ly$\alpha$ EW 
\begin{equation}
    EW_{\rm int} = \frac{8.7\dot{}L_{\rm H\alpha,cor}}{L_{\rm \lambda,1216,cor}}
\end{equation}
where L$_{\rm H\alpha,cor}$ is the dust-corrected and de-magnified H$\alpha$ luminosity, and L$_{\rm \lambda,1216,cor}$ is the dust corrected and de-magnified UV luminosity density extrapolated to 1216 \AA. Figure \ref{fig:intMUV} shows the intrinsic EW as a function of absolute UV magnitude. 

We compare our measured $\rm EW_{int}$ values to those of more massive galaxies at the same redshift. We convert the ionizing photon production efficiency ($\xi_{\rm ion}$) of \citet{Shivaei2018} to $\rm EW_{int}$ by converting $\xi_{\rm ion}$ to $\rm L_{Ly\alpha}/L_{UV}$. In \citet{Shivaei2018} they calculate $\xi_{\rm ion}$ from weighted mean stacks of their galaxies in bins of absolute UV magnitude. They are shown as gold squares in figure \ref{fig:intMUV}. \citet{Shivaei2018} show a mean log$(\xi_{\rm ion}) = 25.36\pm0.06$ for the SMC curve. This corresponds to an intrinsic EW of $23\rm\AA\pm3$ \AA. \citet{Emami2020} have $\xi_{\rm ion} = 25.22\pm0.10$ which corresponds to an intrinsic EW of $22\pm5$ \AA\ for a similar sample. For our faint galaxies our mean intrinsic EW is $67\pm8$ \AA. Therefore, we see a significant increase in the mean intrinsic EW for fainter galaxies. The faintest galaxies in our sample may be skewing the sample toward higher $\rm EW_{int}$. We also show in figure \ref{fig:intMUV} the observed EWs via the colorbar where blue and purple points indicate LAEs. From the plot we observe large intrinsic EWs ($\rm EW_{Ly_\alpha}^{int} > 40$ \AA) facilitate observation of LAEs, but are not a guarantee that a galaxy will be observed as a LAE. This implies that the ionizing output of the stellar population in galaxies has significant bearing on whether a galaxy is observed to be a LAE. However, it is clear that the stellar population alone cannot account for the observability of LAEs.

\subsection{Escape Fraction $f_{\rm esc}$}
\begin{figure}
    \centering
    \includegraphics[height=7cm]{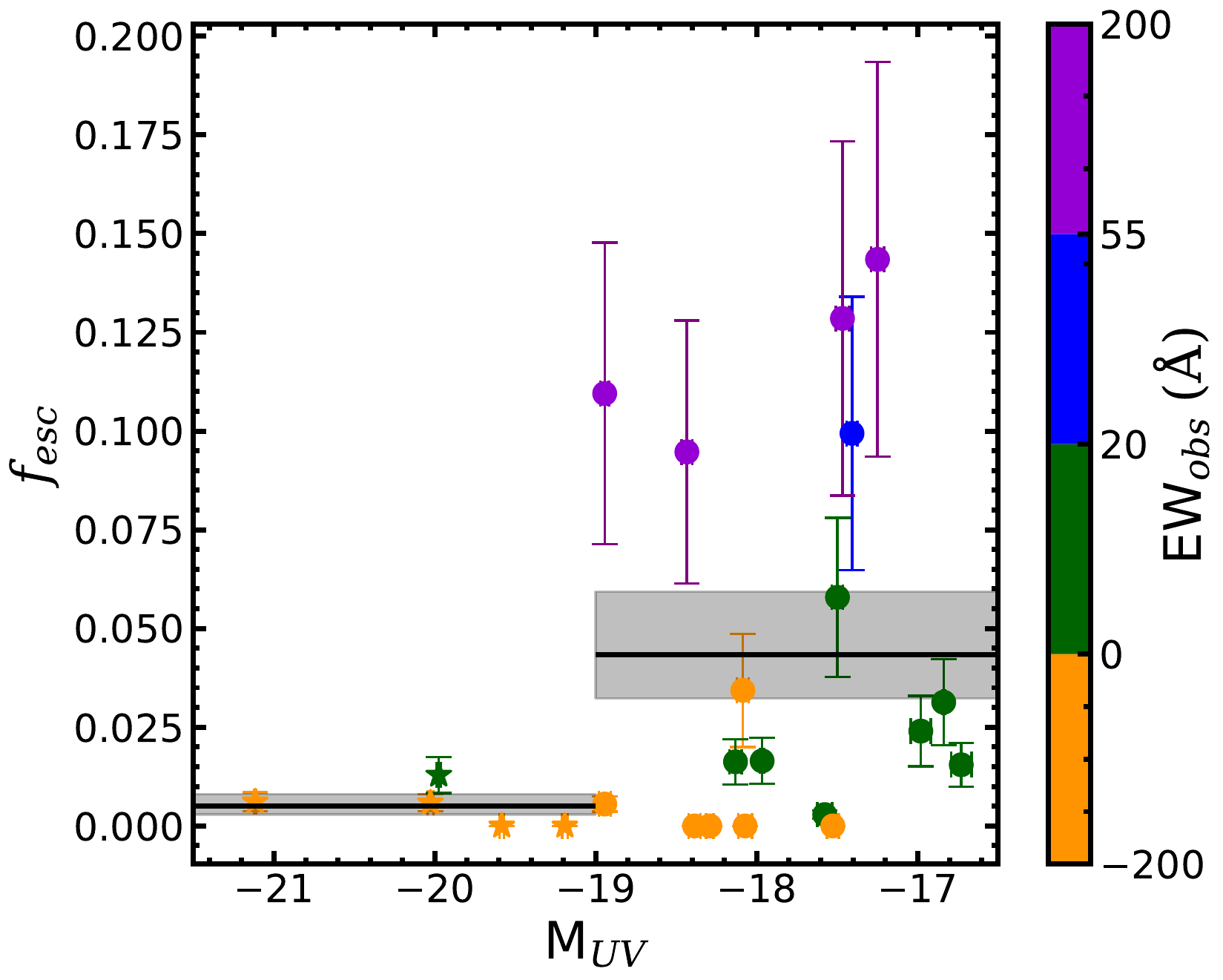}
    \caption{The Ly$\alpha$ escape fraction as a function of absolute UV magnitude (M$_{\rm UV}$). The color coding and markers are the same as in figure \ref{fig:intMUV}. The mean escape fraction of the dwarf galaxies ($4.3^{+1.6}_{-1.1}\%$) and the more massive galaxies ($0.5^{+0.3}_{-0.2}\%$) are displayed with shaded black regions denoting the propagated uncertainty in the individual galaxies. However, the bright galaxy sample has only 5 galaxies and is therefore uncertain.
    \label{fig:fescMUV}}
\end{figure}

The other factor driving the observed distribution in Ly$\alpha$ EWs is whether the photons actually escape from the galaxy. With our MOSFIRE spectra we are able to measure this value directly. For our absorption galaxies in the H$\alpha$ sample we set $f_{\rm esc} = 0$. We show the escape fraction as a function of absolute UV magnitude in figure \ref{fig:fescMUV}. 

We see large scatter in the escape fraction at faint absolute UV magnitudes. It is also the case that no escape fraction less than $\sim 10\%$ is observed as a Ly$\alpha$ emitter. Therefore, it is clear that the escape of ionizing photons plays a crucial role in the observed EW distribution and X$_{\rm LAE}$. 

\begin{figure}
    \centering
    \includegraphics[height=7cm]{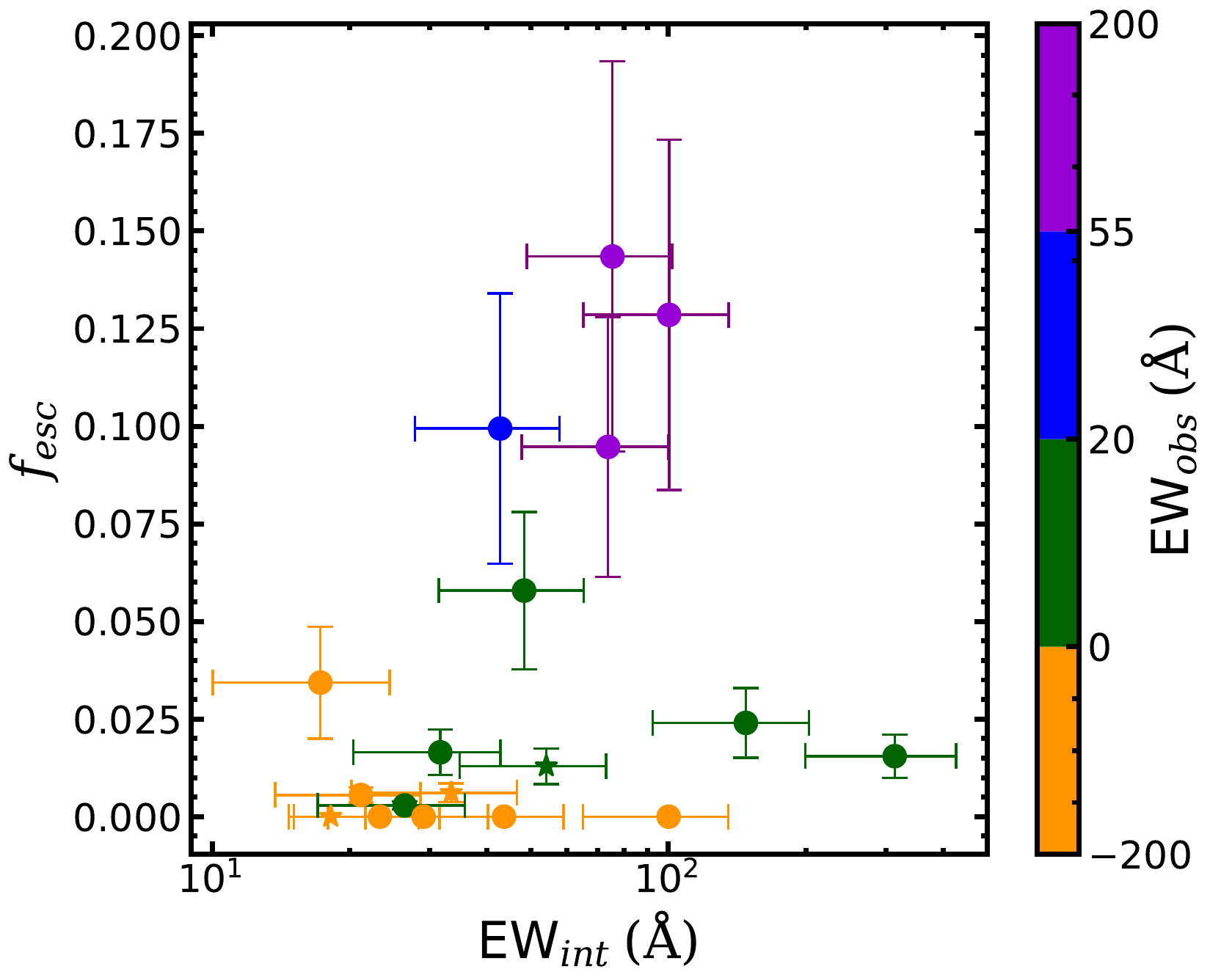}
    \caption{The Ly$\alpha$ escape fraction as a function of \textit{intrinsic} equivalent width. The color coding and markers are the same as in figure \ref{fig:intMUV}. Galaxies with no Ly$\alpha$ in emission are set to $f_{\rm esc}=0$. We observe that only galaxies with intrinsic equivalent widths greater than 40 \AA\ have escape fractions greater than 0.05.}
    \label{fig:fescint}
\end{figure}

To determine if the escape fraction and EW$_{\rm int}$ are correlated, we plot them in figure \ref{fig:fescint}. 

We see that 4/9 of the galaxies with high intrinsic EWs ($\ge 40$ \AA) have large escape fractions ($f_{\rm esc} > 0.05$). No galaxy (0/8) with $\rm EW_{int} < 40$ \AA\ has an escape fraction larger than 0.05. We can set two conditions for LAEs; LAEs have escape fractions greater than 5\% and intrinsic EWs greater than $40$ \AA. We caution that our sample size is small, so this result could be made more robust by increasing the sample size. In order to understand this a little deeper we can look at possible drivers of the escape fraction.

\begin{figure}
     \centering
     \includegraphics[height=7cm]{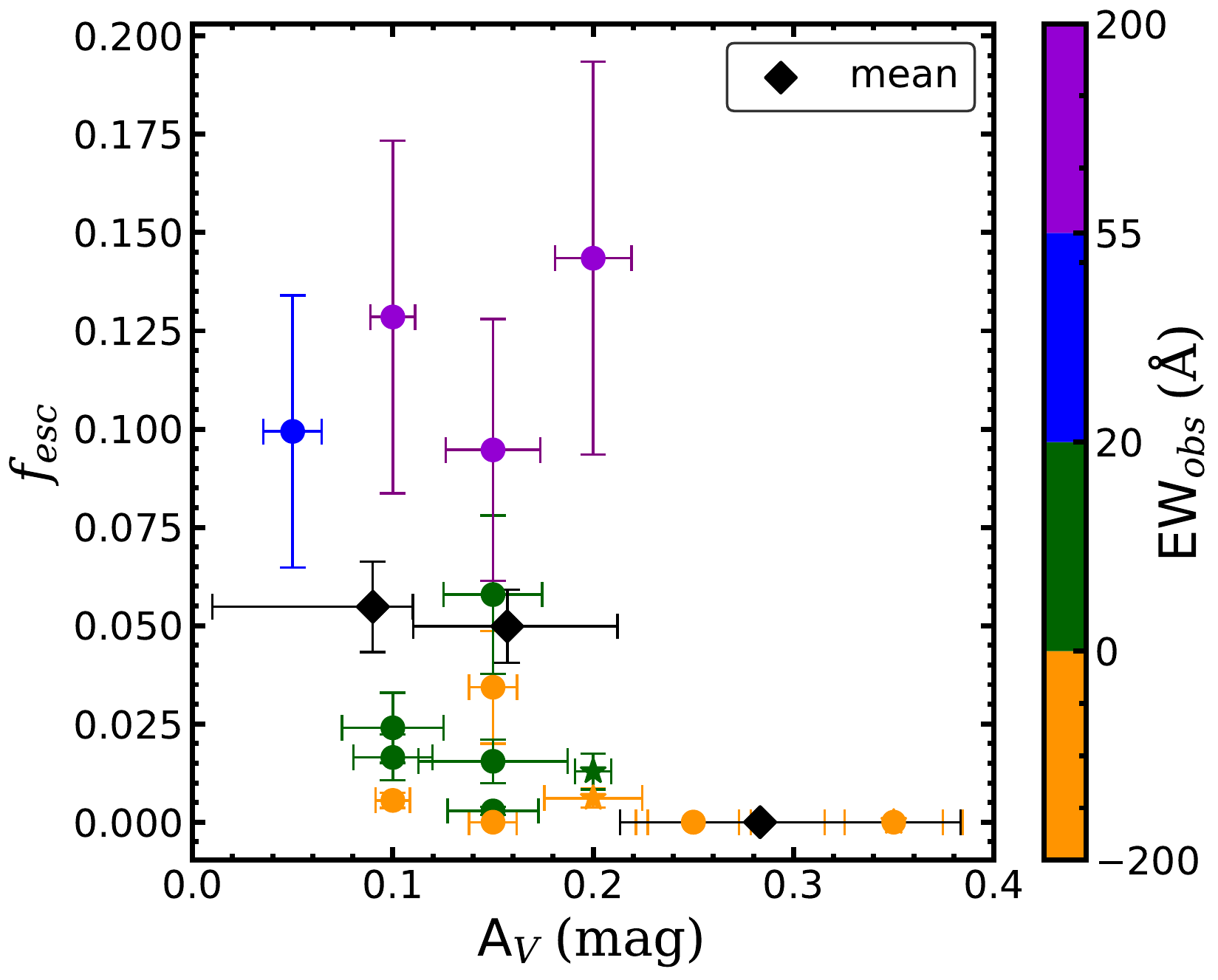}
     \caption{The Ly$\alpha$ escape fraction as a function of dust attenuation. The color coding and markers are the same as in figure \ref{fig:intMUV}. Black diamonds show the mean escape fraction in three bins of $\rm A_V$. The horizontal error bars denote the sizes of the bins and the vertical error bars denote the uncertainty in the mean. A$\rm _V$ is here derived from SED fitting of Hubble photometry and is in discreet steps.}
     \label{fig:fescAV}
 \end{figure}

 Ly$\alpha$ can be heavily attenuated by dust. However, in our sample we have shown that our galaxies are typically not very dusty. Nevertheless we check to see if there is any relation between the dust attenuation estimated from SED fits and the escape fraction. Figure \ref{fig:fescAV} shows the escape fraction as a function of ${\rm A_V}$. We observe an anti-correlation between $f_{\rm esc}$ and ${\rm A_V}$. However, nearly all of our galaxies have $\rm A_V \le 0.2$. With little variance in dust content we aren't able to test this robustly. Nevertheless, we don't observe escape fractions greater than 0.05 for $\rm A_V \ge 0.2$.
 
\begin{figure}
     \centering
     \includegraphics[height=6.5cm]{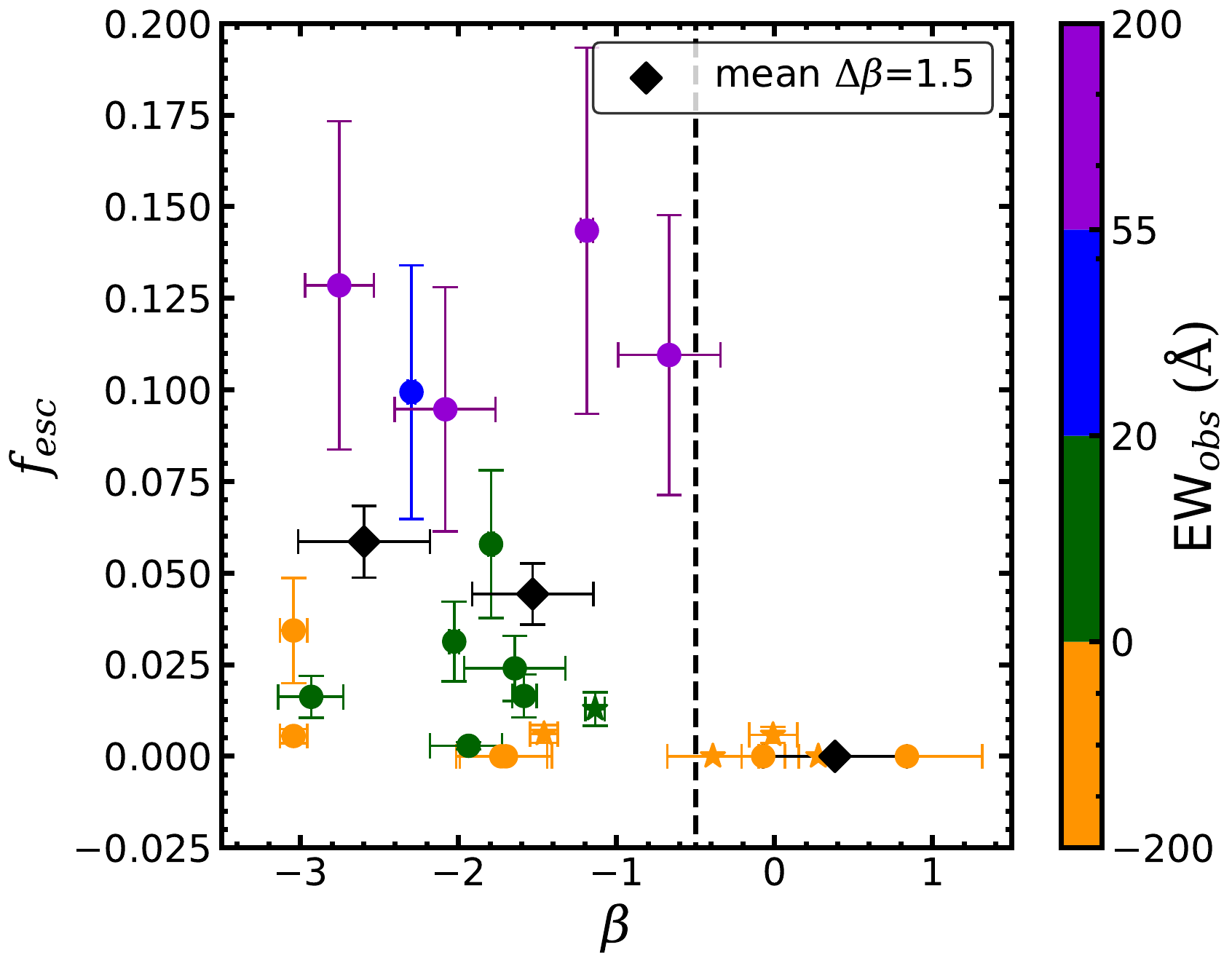}
     \caption{The Ly$\alpha$ escape fraction as a function of UV spectral slope ($\beta$). The color coding and markers are the same as in figure \ref{fig:intMUV}. The black dashed vertical line is set as a limit for LAEs. All LAEs have slopes bluer than -0.5. There is large variance in beta for similar $f_{\rm esc}$, perhaps suggesting bursty star-formation. The black diamonds are mean values of the escape fraction in bins of $\Delta\beta=1.5$. The error bars are calculated by propagating the statistical uncertainty of each datum. One object was excluded due to highly uncertain $\beta$ from low S/N continuum.}
     \label{fig:fescbeta}
 \end{figure}

 The slope of the UV continuum can also be used as an indicator of dust and so we look at possible correlations with UV slope as well. This is shown in figure \ref{fig:fescbeta}. Here we see a mild dependence in that the scatter of escape fractions increases for $\beta < -0.5$. Together, these plots suggest that there is a slight anti-correlation between dust and $f_{\rm esc}$ for faint galaxies. This is consistent with figure \ref{fig:fescAV} which shows an anti-correlation between $f_{\rm esc}$ and A$\rm _V$.
 
 \subsection{Trends in X$_{\rm LAE}$}
 \label{Sec:XLAE}
\begin{figure}
     \centering
     \includegraphics[height=7cm]{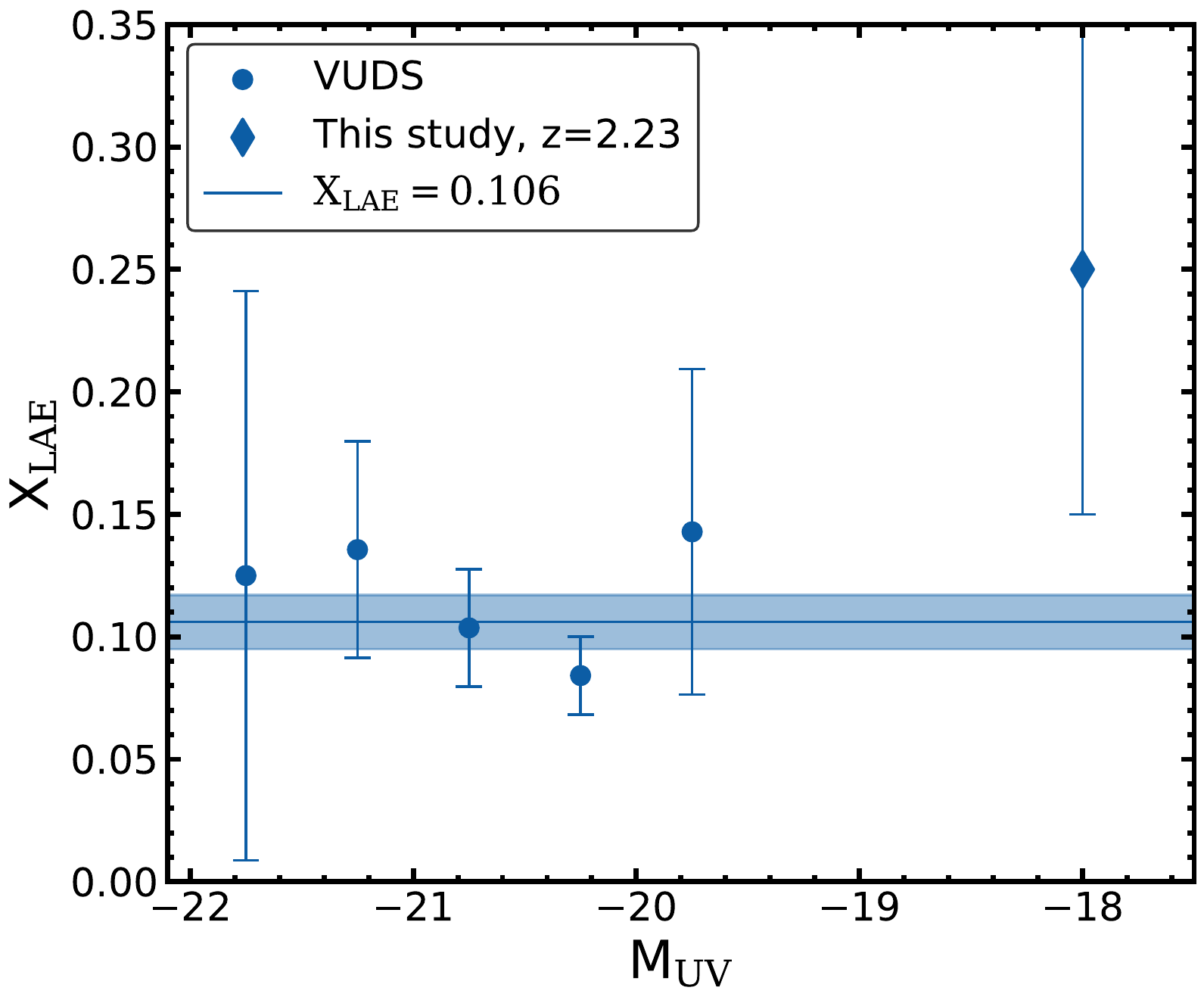}
     \caption{The Ly$\alpha$ emitter fraction, X$_{\rm LAE}$, as a function of M$_{\rm UV}$. Our data set is represented by a blue diamond and the VUDS dataset by blue circles. We show a constant fit line, along with uncertainties in shaded blue, to the VUDS data at X$_{\rm LAE} = 0.106\pm 0.011$ as there is no trend in the brighter M$_{\rm UV}$ sample.}
     \label{fig:XLAE}
 \end{figure}

 The decrement in X$_{\rm LAE}$ from the expected X$_{\rm LAE}$ at high redshift is often used to infer the IGM neutral fraction \citep[i.e.]{Stark2010a,Stark2011a}. However, in order to investigate the number of faint galaxies that contribute to reionization it is helpful to establish trends in X$_{\rm LAE}$ with $\rm M_{UV}$ at lower redshifts where Ly$\alpha$ is observable. To this end we compare with the VUDS sample \citep{LeFevre2015A&A...576A..79L, Hathi2016,Cassata2015b} at higher UV luminosities. Figure \ref{fig:XLAE} shows the VUDS data set in bins of M$_{\rm UV}$ along with our data set. We find no trend in the VUDS data for brighter M$_{\rm UV}$, so we fit a constant line to the VUDS data  set and find X$_{\rm LAE} = 0.106\pm 0.011$. This is $1.4\sigma$ from our result of X$_{\rm LAE} = 0.25^{+0.15}_{-0.10}$. More data between the VUDS faintest M$_{\rm UV}$ and our sample is necessary to determine any trend in X$_{\rm LAE}$ with M$_{\rm UV}$.
 
\begin{figure}
    \centering
    \includegraphics[height=7cm]{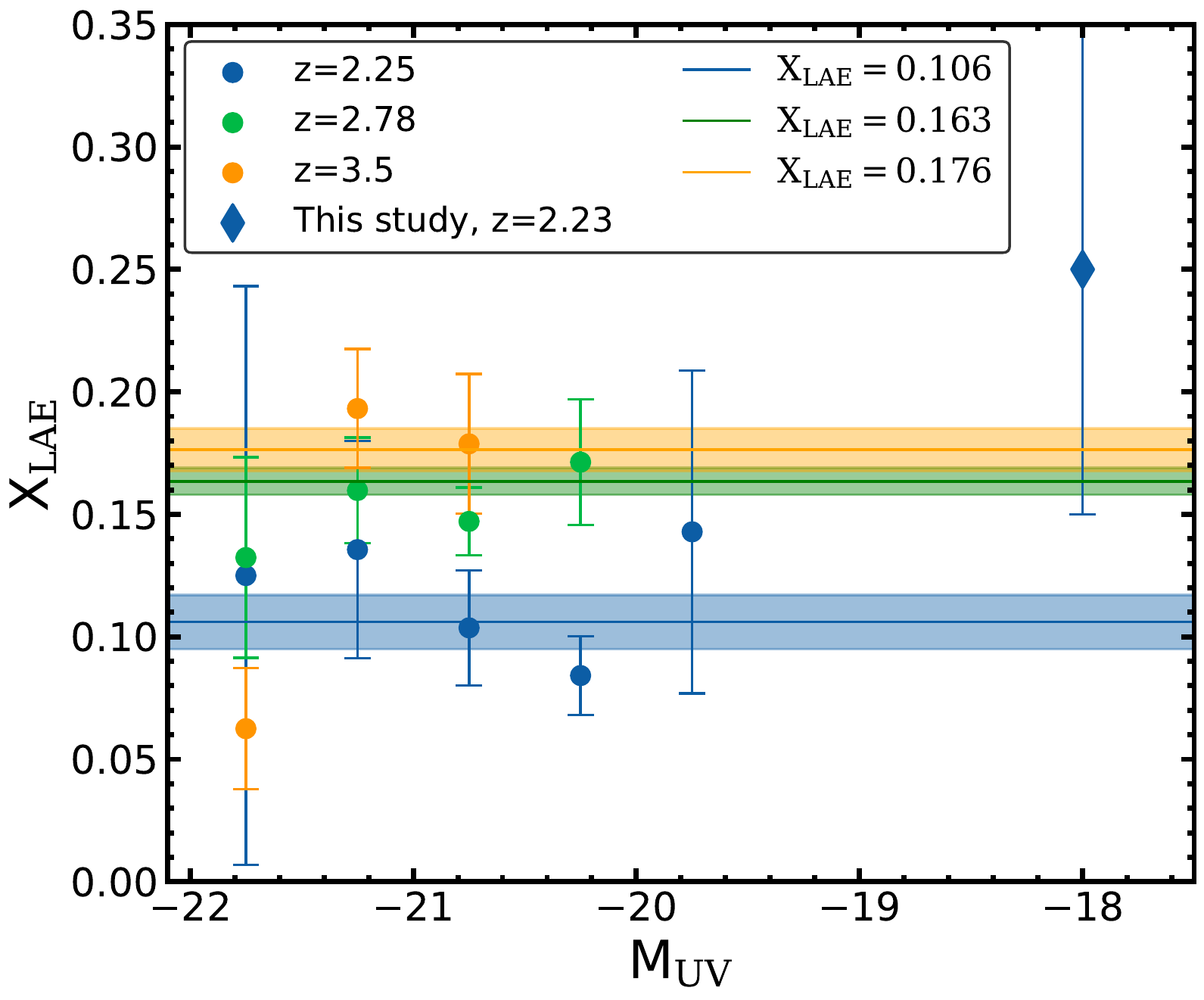}
    \caption{The Ly$\alpha$ emitter fraction, X$_{\rm LAE}$, as a function of absolute UV magnitude and redshift. The lines and shaded regions of corresponding color to the data points are the constant fit curves to the data at each redshift. There are no trends in M$_{\rm UV}$ at any redshift in the VUDS data.}
    \label{fig:ZLAE}
\end{figure}

 We also investigate trends in redshift using the VUDS data set as shown in figure \ref{fig:ZLAE}. We fit each redshift bin with a constant value since we again see no trend with M$_{\rm UV}$ at any redshift. There is a clear trend towards larger X$_{\rm LAE}$ with redshift. The $X_{\rm LAE}$ values of low luminosity galaxies at $z\sim2$ in our sample are consistent with more massive galaxies at $z \ge 3$.

\subsection{Observed Ly$\alpha$ EW}
\begin{figure}
    \centering
    \includegraphics[height=7cm]{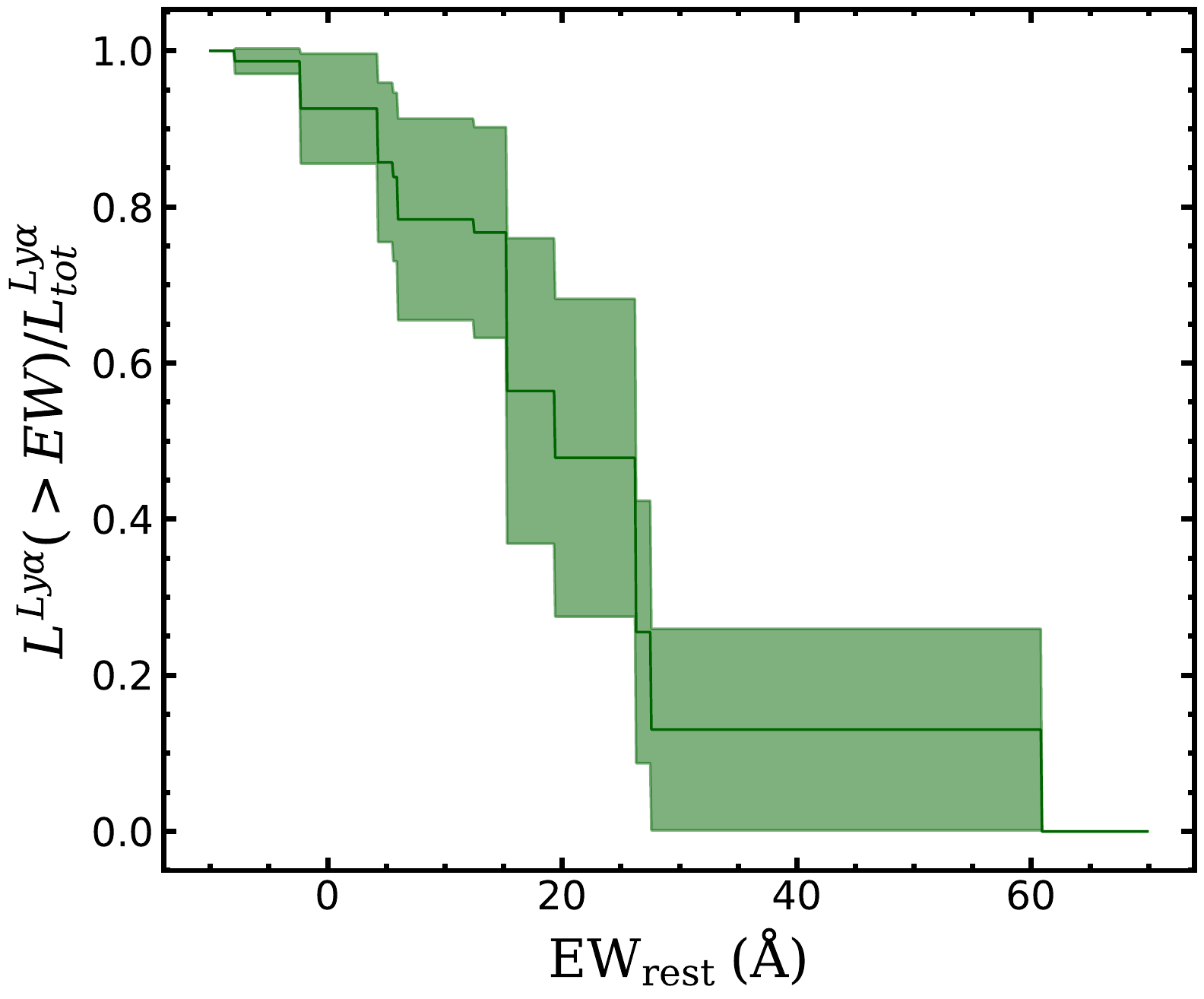}
    \caption{The integrated Ly$\alpha$ luminosity density as a function of EW. The uncertainty is given by the shaded green regions. We find that about 50\% of the integrated luminosity density comes from galaxies with $\rm EW_{Ly\alpha}>20$ \AA}
    \label{fig:sum_lum_dens}
\end{figure}

We investigate which galaxies are emitting most of the Ly$\alpha$ luminosity. Figure \ref{fig:sum_lum_dens} shows that about 50\% of the integrated Ly$\alpha$ luminosity density comes from galaxies with $\rm EW_{Ly\alpha}>20$ \AA. This suggests that LAEs are contributing a large fraction of the Ly$\alpha$ photons. This cumulative distribution can also be calculated from narrow-band surveys for Ly$\alpha$. However, those surveys can not determine the amount of total emissions from low EW galaxies. Here we shos that $\sim10$\% of the Ly$\alpha$ luminosity density EW$_{\rm rest} < 5$\AA\. However, because of the size of our sample, this result is quite uncertain.

 \section{Summary}
 \label{Summary}
 In this paper we have investigated the distribution of Ly$\alpha$ EWs for dwarf galaxies at $z\sim 2$. We selected (via UV continuum) magnified galaxies behind lensing clusters, allowing us to obtain rest-UV spectroscopy for some of the faintest galaxies (M$_{\rm UV} < -17$, $\langle{\rm M}^\star\rangle_{\rm median} = 10^{8.4}\ {\rm M}_\odot$ ) observed at $z\sim2$. We reduce our sample to avoid issues with differential magnification, large slit loss uncertainties,  incompleteness due to faint $H\alpha$, and the observability of Ly$\alpha$ and H$\alpha$ within the bands of Keck/LRIS and Keck/MOSFIRE. We then analyze the EW distribution of the sample and draw the following conclusions:
 \begin{itemize}
 
     \item The observed EW distribution of low UV luminosity galaxies is skewed towards larger EWs than higher luminosity galaxies such as those in the KBSS sample of \citet{Du2021}. Our sample shows a median EW of 5.8 \AA\ whereas the sample of \citet{Du2021} shows a median EW of -6.0 \AA.
     
     \item The fraction of galaxies that are Ly$\alpha$ emitters X$_{\rm LAE}$ is $25^{+15}_{-10}\%,\; 25^{+15}_{-10}\%,\; {\rm and}\; 17^{+13}_{-8}\%$ for $\rm EW>20$ \AA, $\rm EW>25$ \AA, and $\rm EW>55$ \AA\ respectively. These values are greater than X$_{\rm LAE}$ in higher luminosity samples such as \citet{Reddy2008,Hathi2016,Cassata2015b,Du2021}.
     
     \item We investigate possible trends in X$_{\rm LAE}$ with M$_{\rm UV}$ and redshift using the VUDS data set \citep{LeFevre2015A&A...576A..79L, Hathi2016,Cassata2015b}. We find no trend with M$_{\rm UV}$ between $-22\le {\rm M_{UV}} \le -19.5$, but find our X$_{\rm LAE}$ at M$_{\rm UV}\sim-18$ to be $\sim15\%$ greater at 1.4$\sigma$ significance. There is a trend towards larger X$_{\rm LAE}$ with redshift in the VUDS data. Further investigation of dwarf galaxies at higher redshift could show increased X$_{\rm LAE}$ relative to higher mass galaxies as well.
     
     \item We find that the total integrated Ly$\alpha$ luminosity is about 50\% for galaxies with $\rm EW_{Ly\alpha}>20$ \AA, suggesting that LAEs contribute a large fraction of the Ly$\alpha$ photons. However, because of small numbers this value is uncertain.
     
     \item We estimate the intrinsic Ly$\alpha$ EW of galaxies, and show an increase in the mean $\rm EW_{int}$ for faint galaxies ($\rm \langle EW\rangle = 67\pm8\AA$) compared to the brighter sample of \citet[$\rm \langle EW\rangle = 23\pm3\AA$]{Shivaei2018} and the intermediate luminosity sample of \citet[$\rm \langle EW\rangle = 22\pm5\AA$]{Emami2020}. This suggests that younger ages and/or lower metallicities of the stellar populations of dwarf galaxies are increasing the intrinsic EW and contributing to the larger X$_{\rm LAE}$.
     
     \item We investigate the escape fraction of Ly$\alpha$ photons to further understand what is driving the EW distribution. We observe a marginal increase in the mean escape fraction with absolute UV magnitude. We also observe that only galaxies with EW$_{\rm int} > 40\AA$ have $f_{\rm esc}>0.05$. This suggests that only galaxies with higher EW$_{\rm int}$ will be classified as LAEs, not just because of their higher EW$_{\rm int}$, but also because only those galaxies have large escape fractions.
     
     \item We find an anti-correlation between $f_{\rm esc}$ and both  $\rm A_V$ and $\beta$. This implies that low dust-content facilitates the escape of Ly$\alpha$ photons.
     
     \item We observe a global {\it volumetric} escape fraction of $f_{\rm esc} = 4.59^{+2.0}_{-1.4}\%$ in our sample, in good agreement with values found for other faint UV luminosity samples in the literature \citep[i.e.][etc.]{Hayes2010,Weiss2021}.
     
 \end{itemize}
 
We were able to disentangle to some extent two primary drivers of the Ly$\alpha$ EW distribution, namely the ionizing sources (via the intrinsic EW), and the ISM gas/dust content (via the escape fraction and its dependencies). This sample can serve as a baseline with which to compare higher redshift and higher mass samples. Larger sample sizes of dwarf galaxies would serve to solidify some of the results of this work. 
 
\section*{Acknowledgments}
This research made use of {\ttfamily{PypeIt},\footnote{\url{https://pypeit.readthedocs.io/en/latest/}}}
a Python package for semi-automated reduction of astronomical slit-based spectroscopy
\citep{pypeit:joss_pub, pypeit:zenodo}. \\
Some of the data presented herein were obtained at the W. M. Keck Observatory, which is operated as a scientific partnership among the California Institute of Technology, the University of California and the National Aeronautics and Space Administration. The Observatory was made possible by the generous financial support of the W. M. Keck Foundation. \\
Based on observations made with the NASA/ESA \textit{Hubble Space Telescope}, obtained from the Data Archive at the Space Telescope Science Institute, which is operated by the Association of Universities for Research in Astronomy, Inc., under NASA contract NAS5-26555. These observations are associated with programs \#9289, \#11710, \#11802, \#12201, \#12931, \#13389, \#14209.



\section*{Data Availability}

This paper is based on public data from the Hubble Space Telescope as well as from programs 12201, 12931, 13389, 14209.  Spectroscopic data from our survey with the Keck Observatory. These data are available upon request from Christopher Snapp-Kolas or Dr. Brian Siana.



\bibliographystyle{mnras}
\bibliography{Paper1_updated} 





\bsp	
\label{lastpage}
\end{document}